\documentclass[aps,prl,twocolumn,superscriptaddress]{revtex4}

\usepackage{mathrsfs,amssymb,graphics,subfigure,threeparttable}
\usepackage{grap hicx}
\usepackage{amssymb}
\usepackage{enumerate}
\usepackage{amsmath}
\usepackage{fancyhdr}
\usepackage{bm}
\usepackage[squaren]{SIunits}

\begin{document}


\title{Exact phase space matching for staging plasma and traditional accelerator components using longitudinally tailored plasma profiles}


\author{X. L. Xu}
\affiliation{Department of Engineering Physics, Tsinghua University, Beijing 100084, China}
\affiliation{University of California, Los Angeles, California 90095, USA}
\author{Y. P. Wu}
\affiliation{Department of Engineering Physics, Tsinghua University, Beijing 100084, China}
\author{C. J. Zhang}
\affiliation{Department of Engineering Physics, Tsinghua University, Beijing 100084, China}
\author{F. Li}
\affiliation{Department of Engineering Physics, Tsinghua University, Beijing 100084, China}
\author{Y. Wan}
\affiliation{Department of Engineering Physics, Tsinghua University, Beijing 100084, China}
\author{J. F. Hua}
\affiliation{Department of Engineering Physics, Tsinghua University, Beijing 100084, China}
\author{C.-H. Pai}
\affiliation{Department of Engineering Physics, Tsinghua University, Beijing 100084, China}
\author{W. Lu}
\email[]{weilu@tsinghua.edu.cn}
\affiliation{Department of Engineering Physics, Tsinghua University, Beijing 100084, China}
\author{W. An}
\affiliation{University of California, Los Angeles, California 90095, USA}
\author{P. Yu}
\affiliation{University of California, Los Angeles, California 90095, USA}
\author{W. B. Mori}
\affiliation{University of California, Los Angeles, California 90095, USA}
\author{M. J. Hogan}
\affiliation{SLAC National Accelerator Laboratory, Menlo Park, California 94025, USA}
\author{C. Joshi}
\affiliation{University of California, Los Angeles, California 90095, USA}


\date{\today}

\begin{abstract}
Phase space matching between two plasma-accelerator (PA) stages and between a PA and a traditional accelerator component is a critical issue for emittance preservation of beams accelerated by PAs. The drastic differences of the transverse focusing strengths as the beam propagates between different stages and components may lead to a catastrophic emittance growth in the presence of both finite energy spread and lack of proper matching. We propose using the linear focusing forces from nonlinear wakes in longitudinally tailored plasma density profiles to provide exact phase space matching to properly transport the electron beam through two such stages with negligible emittance growth. Theoretical analysis and particle-in-cell simulations show how these structures may work in four different scenarios. Good agreement between theory and simulation is obtained.
\end{abstract}


\maketitle


The invention and continuous development of the charged-particle accelerator in the 20th century have played a very important role in the advancement of modern physics \cite{sessler2007engines}. Even today accelerators such as the Large Hadron Collider \cite{Aad20121} and the Linac Coherent Light Source \cite{emma2010first}, are pushing the frontiers of our knowledge about the origin and complexity of matter. Unfortunately these machines are getting too large and expensive, giving impetus to research on advanced particle acceleration schemes that may lead to a more compact and efficient alternative to the present technology \cite{litos2014high}. One such approach, generally known as plasma-based acceleration has been intensely studied and has made significant recent progress towards both high-gradient and high-efficiency acceleration \cite{leemans2006gev, blumenfeld2007energy, PhysRevLett.103.035002, PhysRevLett.103.215006, PhysRevLett.105.105003, wang2013quasi, PhysRevLett.111.165002, PhysRevLett.113.245002, litos2014high}. However another important challenge in the development of plasma accelerators (PAs) that has only recently been discussed \cite{Antici2012JAP,PhysRevSTAB.16.011302,PhysRevSTAB.15.111303,PhysRevLett.112.035003} and hitherto little explored \cite{PhysRevSTAB.17.054402, PhysRevSTAB.18.041302} is to match the beam out of the plasma into another accelerator component without spoiling the beam's emittance. Emittance preservation is imperative to maintaining the beam's brightness and luminosity for coherent light source and collider applications \cite{Aad20121, emma2010first}.  Therefore, beam matching must be carefully considered when sending the output beam from one PA stage to either a second PA stage or through traditional components used to transport the beam common in accelerators such as focusing magnets. 

In this Letter, we show through both analytical solutions as well as OSIRIS \cite{fonseca2002high} particle-in-cell (PIC) simulations that using plasmas that have longitudinally tailored density profiles as matching sections it is possible to transport the electron beam to/from the PA sections without significant emittance growth using ion channel focusing forces which arise in the nonlinear blowout regime \cite{PhysRevA.44.R6189, PhysRevLett.96.165002, lu2006nonlinearPoP}. We investigate several density profiles,  how to match  the Courant-Snyder (C-S) parameters $\beta$ and $\alpha$ \cite{lee1999accelerator} between the two stages that require beam matching, and exact and adiabatic matching. 

We consider four examples where it will be important to achieve beam-matching between two stages where at least one  stage is a PA. 
The first configuration is the so-called injector-accelerator, where a $\sim100$ MeV class electron beam produced by a short, high-density injector stage is further accelerated to $\sim$ GeV level using a second low-density accelerator stage \cite{gonsalves2011tunable, PhysRevLett.107.035001, PhysRevLett.111.165002}. The second example is the external injection scheme where a high-quality, relativistic electron bunch is first generated using an RF accelerator and then injected into a PA \cite{PhysRevLett.70.37, clayton1994acceleration, everett1994trapped, Stragier2011externalinjection, Rossi201460}. The third example concerns the proposed PA driven light source \cite{schlenvoigt2007compact, fuchs2009laser, cipiccia2011gamma}, where a high-quality electron beam needs to be coupled from the plasma wake to an undulator. The last configuration is for the recently developed collider concepts based on linking together many PAs \cite{leemans2009laser, adli2013beam}.
Each stage (with a separate driver) provides about 10 GeV gain. The distance between the successive stages needs to be sufficient (on the order of one meter) to place beam transport components for coupling the fresh driver to each stage. In the latter three cases a magnetic focusing optic will be needed to couple the beam from one stage into/from the PA.

In the above scenarios, the beam exiting one stage needs to be coupled into the next stage that may have a drastically different field-focusing strength. In traditional accelerators, solenoids and quadrupoles are typically combined to guide the transverse motion of the particles between the stages. However, due to ultra-high focusing gradient in the nonlinear plasma wake ($ G [\mega\tesla/\meter] \equiv F_r/ecr \approx 3.01 n_p [10^{17}\centi\meter^{-3}]$), state of the art quadrupoles ($G\sim 10^3~\tesla\per\meter$) \cite{PhysRevSTAB.10.082401, PhysRevSTAB.15.070703} are not strong enough to confine the transverse motion of the particles between the stages. Here $F_r$ is the transverse focusing force in the direction $r$ and $n_p$ is the plasma density. As a result, beams will experience orders of magnitude transverse size variation when propagating between the PA and the conventional focusing optic, and the particles' transverse motion will become very sensitive to the energy spread of the bunch, i.e., particles with different energy will undergo transverse betatron oscillations with different betatron phases, leading to a catastrophic emittance growth \cite{Antici2012JAP, PhysRevSTAB.15.111303, PhysRevSTAB.16.011302, PhysRevLett.112.035003}. 

\begin{figure}[bp]
\includegraphics[width=0.5\textwidth]{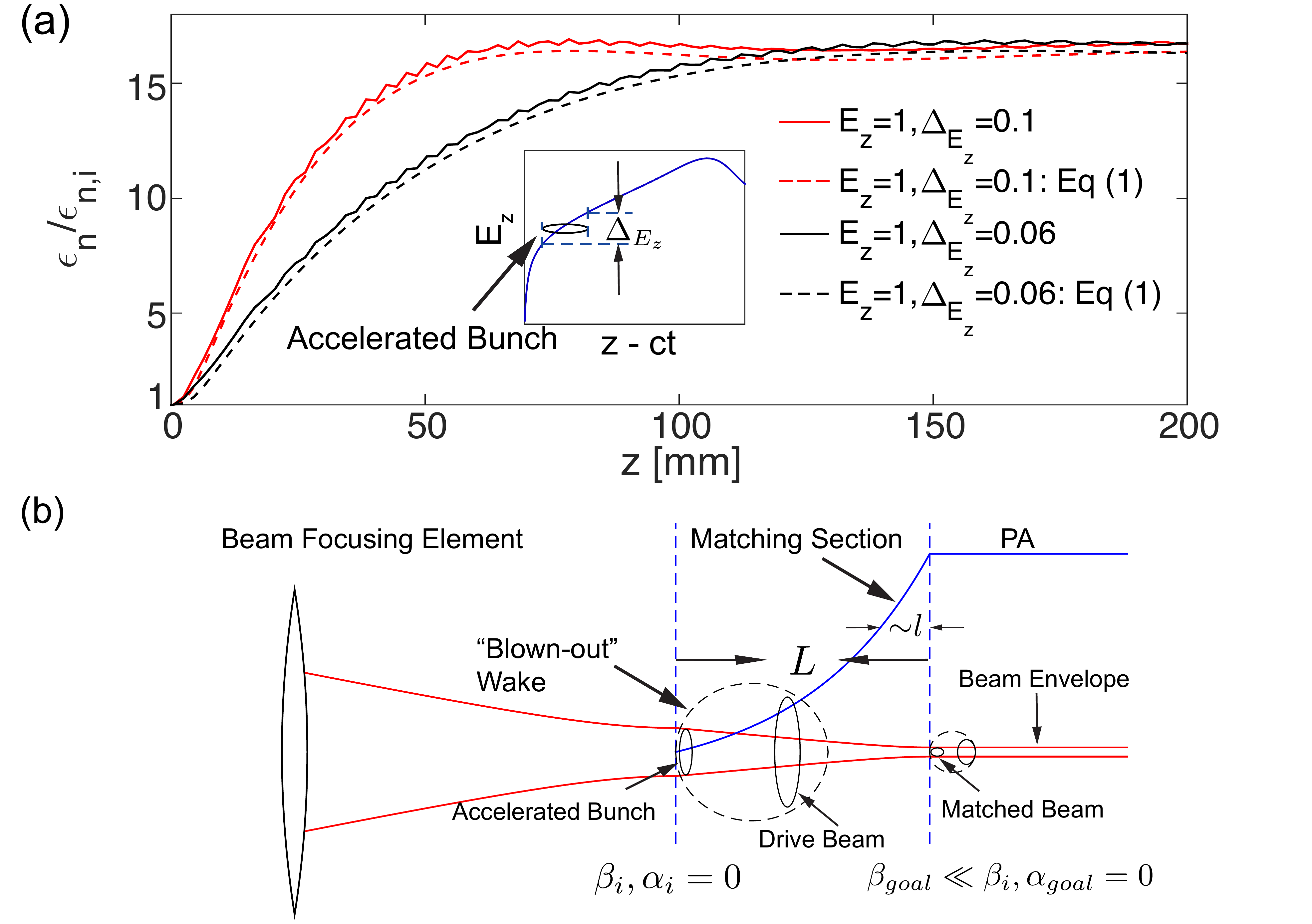}
\caption{\label{fig:} (a) The emittance evolution of a 100 MeV electron beam from a high density plasma injector as it propagates in a low density plasma accelerator. The emittance evolution for two different values of $\Delta_{E_z}$ and a normalized $E_ze/mc\omega_p=1$. The inset shows the relative position of the bunch within the nonlinear accelerating cavity. (b) The concept of matching using a longitudinally tailored plasma profile. The beam to be accelerated in a PA is focused at the entrance of a plasma density ramp for matching and injected into a fully "blow-out" wake produced by either a laser pulse or an electron bunch (driver bunch).}
\end{figure}

The transverse normalized emittance, which is a figure of merit for the beam quality, is defined as $
\epsilon_n=\frac{1}{mc}\sqrt{ \left\langle x^2 \right\rangle \left\langle p_x^2 \right\rangle - \left\langle x p_x \right\rangle ^2} $, where $\left\langle \right\rangle$ represents an ensemble average over the beam distribution, $x$ is the transverse position and $p_x$ is the transverse momentum. The phase space distribution is described by the C-S parameters $\beta,\alpha$ and $\gamma$ \cite{lee1999accelerator} where $\beta = \left\langle x^2 \right\rangle / \epsilon, \alpha = \left\langle x x' \right \rangle / \epsilon, \gamma = \left\langle x'^2 \right\rangle / \epsilon$, where $x'={\mathrm{d}x}/{\mathrm{d}z}=p_x/p_z$ is the slope of the particle trajectory, $\epsilon = \sqrt{ \left\langle x^2 \right\rangle \left\langle x'^2 \right\rangle - \left\langle x x' \right\rangle ^2}$ is the geometric emittance, $\beta$ is a measure of the beam size, $\alpha$ represents the correlation between $x$ and $x'$ (e.g., at beam waist $\alpha=0$), and $\gamma$ is a measure of the spread in the particle slopes. The C-S parameters satisfy the relationship  $\beta \gamma = 1+\alpha^2$. In typical cases, the C-S parameters of a matched electron beam in the PAs are determined by the field structure inside the nonlinear wake as $\beta_p=\sqrt{2 \left \langle \gamma_b \right\rangle }k_p^{-1}, \alpha_p = 0$, where $ \left \langle \gamma_b \right\rangle$ is the average value of the relativistic factor of the beam.

It is straightforward to obtain the emittance evolution when a relativistic beam drifts in free space as $\epsilon_n \left( z \right)= \left\langle p_z \right \rangle \epsilon \sqrt{\hat{\sigma}_{\gamma_b}^2 \left[ \left( \gamma_i z - \alpha_i \right)^2 + 1\right] + 1 } $ \cite{Antici2012JAP}\cite{PhysRevSTAB.16.011302}, where $\hat{\sigma}_{\gamma_b} = \sqrt{  \left\langle p_z^2 \right\rangle - \left \langle p_z \right \rangle ^2 } /  \left \langle p_z \right \rangle$ is the relative energy spread of the beam, and the geometric emittance $\epsilon$ remains constant in free space. Here subscript `i' refers to the input or initial quantity. When the relativistic beam propagates in focusing elements, the emittance evolution is determined by the detailed configurations of the quadrupoles or the field structure in the plasma wake. For the simple case where a linear focusing force $F_r$ that is constant in $z$ is present, the emittance grows and finally saturates when the beam is not matched and there is any initial or induced energy spread. Now we consider the situation shown in Fig. 1(a) where both $F_r$ and accelerating field $E_z$ are present. Here an electron bunch of $ \left \langle \gamma_{b,i} \right\rangle=200$ with an initial energy spread $\hat{\sigma}_{\gamma_b}=0.01$ is produced in a $10^{19}~\centi\meter^{-3}$ injector stage ($\beta_i=33.7~\micro\meter, \alpha_i=0$). It then propagates 0.5 mm in vacuum ($\beta_v\approx220\beta_i, \alpha_v\approx15$) before entering a lower density ($10^{17}~\centi\meter^{-3}$) acceleration stage with no attempt made to match the beam between the two stages. 
Further energy spread is induced by the acceleration gradient that varies uniformly between $\left[E_z-{\Delta_{E_z}}/{2}, E_z+{\Delta_{E_z}}/{2} \right]$. Numerical results for the evolution of the emittance are plotted as solid line in Fig. 1(a) for two different values of $\Delta_{E_z}$. Catastropic emittance growth by more than a factor of 15 is seen.

It is also possible to obtain an analytical expression for the projected emittance. Following the derivation in Ref. \cite{PhysRevLett.112.035003} for cases where all particles are initialized at the same $z_i$ leads to 
\begin{align}
\epsilon_n  =  \epsilon_{n,sat} \sqrt{1 - \frac{(\gamma_i \beta_F + \beta_i / \beta_F)^2-4}{(\gamma_i \beta_F + \beta_i / \beta_F)^2} \left(\frac{\mathrm{sin}\Delta\Phi}{\Delta\Phi}\right)^2}
\end{align}
where $\epsilon_{n,sat} \approx \epsilon_{n,i}\left( \gamma_i \beta_F + \beta_i / \beta_F\right)/2$ \cite{PhysRevSTAB.15.111303} and $\beta_F=\sqrt{ \left \langle \gamma_b \right\rangle  m c / Ge}$ is the average beta function of the beam within the focusing element. Here $\Phi$ is the electron betatron phase and is assumed to be uniformly distributed over $\Delta\Phi$. If the particles are not being accelerated, $\Phi=\Phi_i+z/\sqrt{  \gamma_b m c / Ge}$, while if the particles are being accelerated then $\Phi=\Phi_i+(\sqrt{2\gamma_b}-\sqrt{2\gamma_{b,i}})/E_z$ and $\beta_F$ in Eq. (1) corresponds to the value when the beam enters the focusing element. The emittance growth from Eq. (1) using the values for $\gamma$ and $\beta$ at the end of the vacuum section as the initial values is plotted as dashed lines in Fig. 1(a) and excellent agreement with the numerical results can be seen. 

As seen from the above example, the emittance of the beam will grow quickly as the beam propagates if it has a finite energy spread and is not matched between the focusing elements. However by using a plasma that has a specific longitudinal density profile (matching section) one can guide the beam through the two stages with negligible emittance growth. The proposed density-profile matches the initial $\beta_i$ of the bunch to the $\beta_{goal}$ of the PA or the external focusing elements by providing a continuously varying focusing force to transport the bunch from its waist ($\alpha_i=0$) at the exit of the first stage to another waist ($\alpha_{goal}=0$) at the end of the matching section [see Fig. 1 (b)]. In all four cases mentioned earlier, it is possible to match the beam from one stage into another using this technique while preserving the beam emittance. The use of  tailored focusing profiles and linear wakes to couple the particle beam into/from a PA stage has been previously suggested in the adiabatic limit \cite{PhysRevSTAB.17.054402, PhysRevSTAB.18.041302}. However,  linear wakes unlike nonlinear wakes have nonlinear focusing forces, axial dependent focusing forces, and focusing forces which are altered by beam loading \cite{PhysRevLett.101.145002}. Therefore, we consider both the plasma accelerator and the tailored density ramp to be in the nonlinear blowout regime \cite{PhysRevLett.96.165002, lu2006nonlinearPoP}. Note that this theory can also be used in the adiabatic limit and as an estimate for matching electron or positron beams from/to plasma accelerator using linear wakes.

We start with the equation for the transverse motion of a single electron in the blowout regime (linear focusing force) in a density ramp,
\begin{align}
\frac{\mathrm{d}^2 x}{\mathrm{d}z^2} + K(z) x=0
\label{eq:motion equation1}
\end{align}
where $K(z)=n_{p0}f(z)e^2/\left(2 \gamma_b  mc^2\epsilon_0\right) = f(z) \beta_{p0}^2$, $n_{p0}$ is the peak density at the beginning of the matching plasma, and $f(z)$ is the normalized plasma density profile. We also assume that the beam is in a region where there is negligible acceleration in the matching section. We can normalize all the lengths to $\beta_{p0}$, then Eq. (\ref{eq:motion equation1}) can be expressed as $\mathrm{d}^2 \hat{x}/\mathrm{d}\hat{z}^2 + f(\hat{z})\hat{x}=0$, where $\hat{x}=x/\beta_{p0}, \hat{z}=z/\beta_{p0}$. We have found solutions to Eq. (\ref{eq:motion equation1}) for five different density profiles used in Fig. 2. As we will show, the profile with the best matching properties is $f(\hat{z}) =  \hat{l}^2 / (\hat{z}+\hat{l})^2$, so we analyze this case in more detail. For this profile (when $\hat{l}>1/2$) the solution to Eq. (\ref{eq:motion equation1}) is,
\begin{align}
\hat{x}&=c_1 \sqrt{\xi} \mathrm{cos}\Phi+c_2 \sqrt{\xi} \mathrm{sin}\Phi \\
\hat{x}'&= \frac{c_1}{\sqrt{\xi}} \left( \frac{\mathrm{cos}\Phi}{2} - s\mathrm{sin}\Phi\right)  + \frac{c_2}{\sqrt{\xi}}\left( \frac{\mathrm{sin}\Phi}{2} + s\mathrm{cos}\Phi\right)
\end{align}
where $\xi=\hat{z} + \hat{l}$, $s=\sqrt{\hat{l}^2-1/4}$, $\Phi=s\mathrm{ln}\xi$ is the betatron phase advance of the electron, and $c_1, c_2$ are constants determined by the initial conditions for $\hat{x}$ and $\hat{x}'$. Eqs. (3) and (4) can then be used to obtain the mapping  $\begin{pmatrix}
\hat{x}\\
\hat{x}'
\end{pmatrix}=M\left( \hat{z} | 0\right)\begin{pmatrix}
\hat{x}_i\\
\hat{x}'_i
\end{pmatrix}$, which defines the transport matrix. The elements of the transport matrix can be used to express the C-S parameters at $\hat{z}$  in terms of their initial values \cite{lee1999accelerator} 
\begin{align}
\label{eq:paras1} \hat{\beta}_{\hat{l}} (\hat{z}) &= M_{11}^2\hat{\beta}_i -2 M_{11}M_{12}\alpha_i+ M_{12}^2 \hat{\gamma}_i \\
\label{eq:paras2} \alpha_{\hat{l}} (\hat{z}) &= -M_{11}M_{21}\hat{\beta}_i + \left( M_{11}M_{22}+M_{12}M_{21} \right) \alpha_i \nonumber \\
&- M_{12}M_{22}\hat{\gamma}_i
\end{align}

\begin{figure}[bp]
\includegraphics[width=0.5\textwidth]{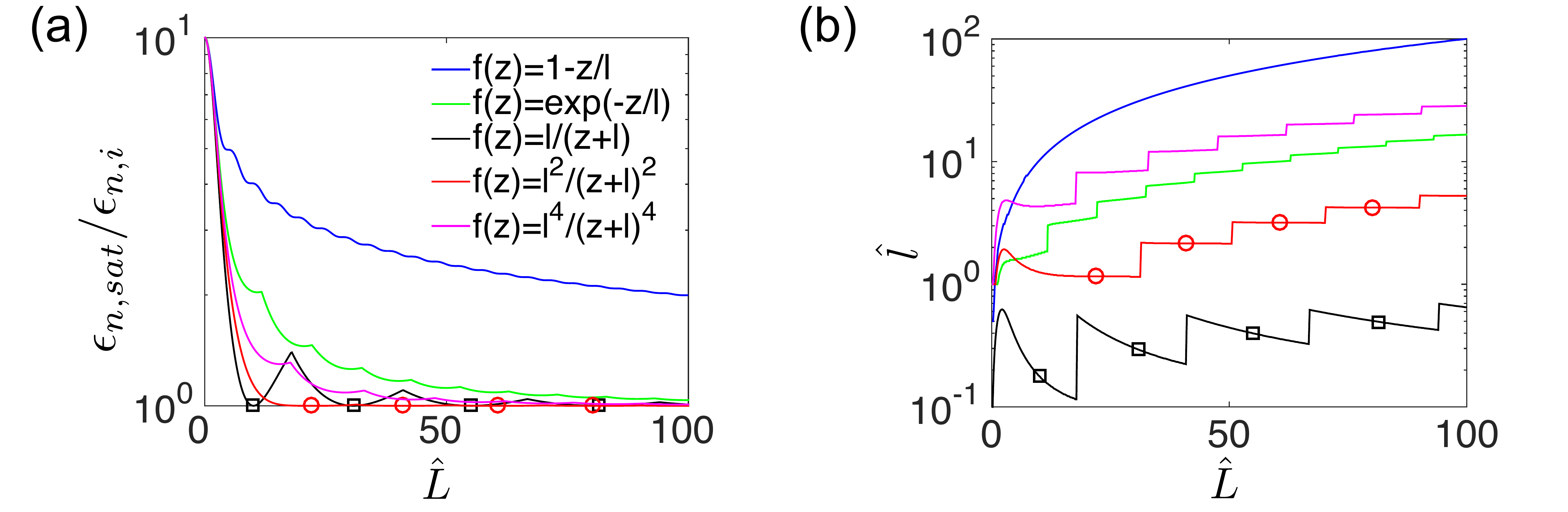}
\caption{\label{fig: minimum emittance} The performance of the matching plasma with different density profiles. For each $\hat{L}$, $\hat{l}$ is scanned to find the optimized value. The parameters: $\hat{\beta}_i=1, \alpha_i=0, \hat{\beta}_{goal}=20, \alpha_{goal}=0$; case of transition from a PA to a magnetic optics. 
}
\end{figure}

A given matching section has a length $z_{max} \equiv \hat{L}$. For a selected $\hat{L}$ the output $\beta$ and $\alpha$ will depend on $\hat{l}$. There will be an optimum $\hat{l}$ such that the emittance growth is minimized within the target section which has a beta function, $\beta_{goal}$. To obtain the optimum $\hat{l}$ we minimize $\epsilon_{n,sat}/\epsilon_{n,i}= \left[{\gamma_{\hat{l}}(\hat{z})\beta_{goal} + \beta_{\hat{l}}(\hat{z})/\beta_{goal}}\right]/{2}$ for fixed $\hat{z} =\hat{L}$, and $\beta_{goal}$. Here subscript `goal' refers to the final desired quantity. We use Eqs. (5) and (6) to obtain $\hat{\beta}_{\hat{l}}(\hat{z})$ and $\hat{\alpha}_{\hat{l}}(\hat{z})$ for given initial C-S parameters. 

In Fig. 2 (a) we plot the optimum $\epsilon_{n,sat}/\epsilon_{n,i}~vs.~\hat{L}$ for a particular $\beta_{goal}$, $\beta_i$, and $\alpha_i$ for each of the five density profiles. By using mathematica \cite{mathematica}, solutions to Eq.(2) and the appropriate transport matrix can be found for the additional density profiles to generate the curves. In Fig. 2(b) we present the optimum $\hat{l}$ as a function of $\hat{L}$.  The red ($f(\hat{z})=\hat{l}/(\hat{z}+\hat{l})$) and green ($f(\hat{z})=\hat{l}^2/(\hat{z}+\hat{l})^2$) curves are of particular interest because for discrete values of $\hat{L}$ an optimum $\hat{l}$ can be found which provides exact matching conditions. These are shown as squares and circles. Furthermore, the red curve has nearly perfect matching for all $\hat{L} > 10$. For the other density profiles (including $f(\hat{z})=\hat{l}^4/(\hat{z}+\hat{l})^4$) the beam becomes adiabatically matched, i.e., $\epsilon_{n,sat}/\epsilon_{n,i}$ approaches unity as $\hat{L}$ increases.

When matching from a positive phase space ellipse (i.e., $\alpha < 0$) to another positive phase space ellipse, for the  $f(\hat{z})=\hat{l}^2/(\hat{z}+\hat{l})^2$ density profile, the parameters for exact matching can found analytically,
\begin{align}
l = \beta_{p0}\sqrt{\left[\frac{ (N+1)\pi}{\mathrm{ln}{\beta_{goal}}/{\beta_i}}\right]^2+\frac{1}{4}}, \frac{L}{l}=\left( \frac{\beta_{goal}}{\beta_i}-1\right)
\end{align}
where $N=0, 1, 2, \ldots$. For the profile with $f(\hat{z})=\hat{l}/(\hat{z}+\hat{l})$, it is difficult to give an analytical solution of the parameters for exact matching, however for when $\hat{l} \ll 1$ we have found the fitting formulas give near perfect matching
\begin{align}
l &\approx \frac{1.7+N}{2} \beta_{p0}\left(\frac{\beta_{goal}}{\beta_i}\right)^{-0.55}\nonumber \\
\frac{L}{l} &\approx \left[ 0.71 + \frac{(0.75+N)\pi}{2} \frac{\beta_{p0}}{l}  \right]^2-1 
\end{align}

We have considered cases where $\beta_{goal} > \beta_i$ so that a density downramp is needed. We note that there is symmetry between the upramp and downramp cases. For the upramp case,  $\beta_{goal}/\beta_i$ in Eqs. (7) and (8) should be replaced with $\beta_{i}/\beta_{goal}$.

\begin{figure}[bp]
\includegraphics[width=0.5\textwidth]{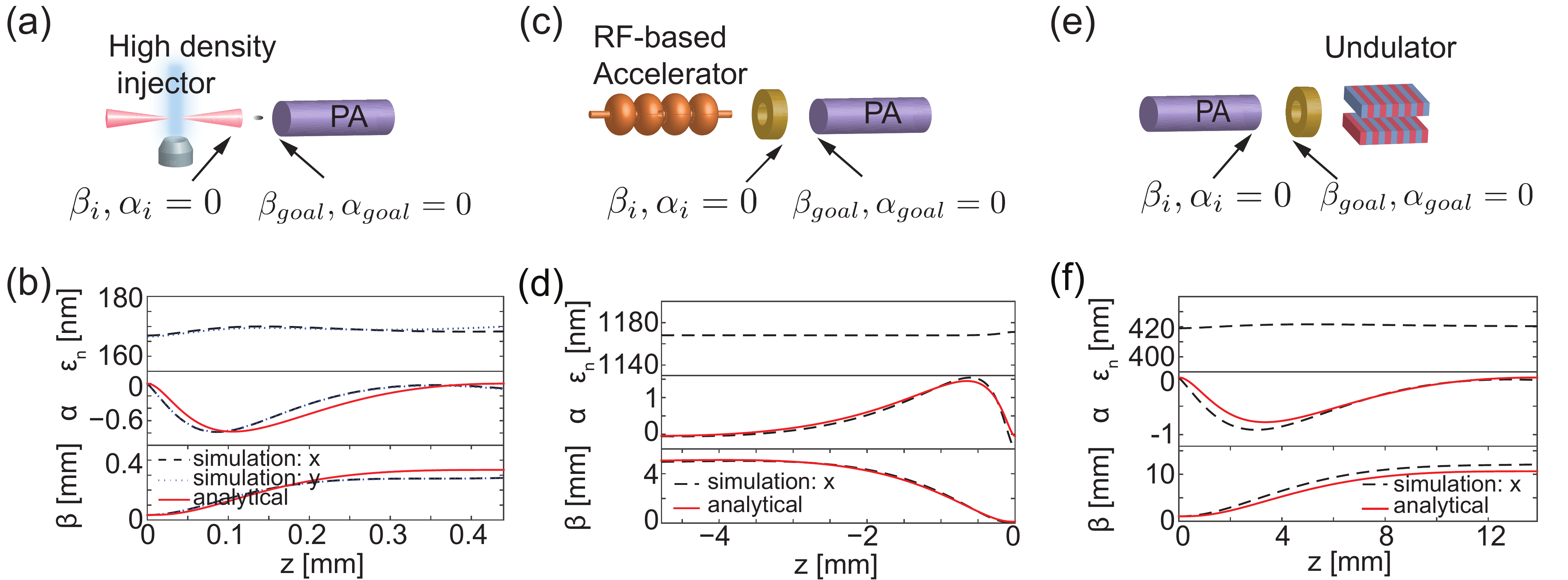}
\caption{\label{fig:inj2acc} 
Schematic of  staging (a) a high density plasma injector and a low density PA, (c) an RF-based injector and a PA using a magnetic optic, and (e) a PA and an undulator using a magnetic optic.   In (b), (d), and (f) the evolution of $\epsilon_n$, $\beta$  and $\alpha$ of the electron beam in the matching section for scenarios (a), (c), and (e) respectively.  For case (b), the driver laser is focused at $z=-0.04~\milli\meter$ , with $a_0=4, w_0= 10~\micro\meter, \tau_{FWHM}= 15~\femto\second$; at $z=0~\milli\meter$ the electron beam is initialized with  $\sigma_{x,y} = 0.17~\micro\meter, \tau_{FWHM} = 5~\femto\second$, and $n_b = 10^{20}~\centi\meter^{-3}$; and between $z=0~\milli\meter$ and $z=0.44~\milli\meter$ the beam parameters vary from  $ \left \langle \gamma_b \right\rangle  = 200$ to $192$ and $\hat{\sigma}_E =0.1$ to $0.105$. For case (d), the driver laser is focused at $z=0 ~\milli\meter$ , with $a_0=3, w_0= 58~\micro\meter, \tau_{FWHM}=100~\femto\second$; at $z=-4.8~\milli\meter$ the electron beam is initialized with  $\sigma_{x} = 10.9~\micro\meter, \tau_{FWHM} = 25~\femto\second$, and $n_b = 10^{16}~\centi\meter^{-3}$; and between $z=-4.8~\milli\meter$ and $z=0~\milli\meter$ the beam parameters vary from  $ \left \langle \gamma_b \right\rangle  50$ to $44.8$ and $\hat{\sigma}_E =0.02$ to $0.0225$. And for case (f), the driver laser is focused at $z=0~\milli\meter$ , with $a_0=3, w_0= 58~\micro\meter, \tau_{FWHM}= 100~\femto\second$; at $z=0~\milli\meter$ the electron beam is initialized with  $\sigma_{x} = 0.34~\micro\meter, \tau_{FWHM} = 25~\femto\second$, and $n_b = 10^{18}~\centi\meter^{-3}$; and between $z=0~\milli\meter$ and $z=13.9~\milli\meter$ the beam parameters vary from  $ \left \langle \gamma_b \right\rangle  = 4000$ to $3966.6$ and $\hat{\sigma}_E =0.05$ to $0.0506$.
} 
\end{figure}

Next, we verify that plasma matching sections can provide nearly perfect matching using fully self-consistent PIC simulations using the code OSIRIS in 3D (or 2D) Cartesian geometry using a moving window \cite{fonseca2002high}. We consider the three examples schematically shown in Fig. 3(a), (c) and (e). In each case we use longitudinally tailored plasma density structures with the ideal density profile $f(z)=l^2/(z+l)^2$ to match the electron beam between stages. We use a laser driver with $\lambda_0=800 \nano\meter$ and define the $z$-axis to be the propagating direction of the drive laser and defined $z=0$ at the peak of the density. The separation of the peak intensity of the laser and density of the electron beam is  $\sim 6c/\omega_p$ in each case. Parameters specific to each simulation are given in the figure caption.

First, we consider matching an electron beam from a high density plasma injector into a low density PA as shown in Fig. 3(a) - the case considered in Fig. 1(a) except now the drift space is replaced by a matching plasma section with final $\beta_{goal} = 337~\micro\meter,\alpha_{goal}=0$. The plasma section has $l\approx 49~\micro\meter, L\approx 440~\micro\meter$, and $N=0$. The 3D simulation has a dimension of $180k_0^{-1}\times 240 k_0^{-1}\times240k_0^{-1}$ with $900\times 1200\times1200$ cells in the $x,y$ and $z$ directions respectively, where $k_0$ is the wavenumber of the driver laser. As can be seen in Fig. 3(b), the matching section aids in preserving the emittance of the electron bunch at its initial level without appreciable growth as opposed to the case shown in Fig. 1(a) and excellent agreement between theory and simulation is found. 

In the second case, we consider matching an electron bunch (from an external accelerator) that is focused at the beginning of the rising density matching section to the PA [see Fig. 3 (c)]. We use 2D simulations with a moving window of $1600k_0^{-1}\times 3000 k_0^{-1}$ with $8000\times 1500$ cells in the $x$ and $z$ directions respectively. The electron beam with $\ \left \langle \gamma_b \right\rangle  =50, \beta_{i} = 5~\milli\meter, \alpha_i = 0$ needs to be exactly matched to $\beta_{goal} = 0.12~\milli\meter,\alpha_{goal}=0$. We use $l\approx 0.12~\milli\meter, L\approx 4.8~\milli\meter$, and $N=0$. Once again the initial beam emittance (1165 nm) is preserved as the beam is transported to the PA and excellent agreement between theory and simulation is found. 

In the third case [Fig. 3 (e)] we consider coupling the electron bunch from the PA via the matching section into a conventional focusing optic so that it can be injected into an undulator.  We use 2D simulations with a moving window of $1600k_0^{-1}\times 3000 k_0^{-1}$ with $8000\times 3000$ cells in the $x$ and $z$ directions respectively. We simulate matching an electron beam leaving a plasma with $ \left \langle \gamma_b \right\rangle  =4000, \beta_{i} = 1.06~\milli\meter, \alpha_i = 0$ out of a matching plasma ($l\approx 1.5~\milli\meter, L\approx 14~\milli\meter$, and $N=0$) into a conventional optic with $\beta_{goal} = 10.6~\milli\meter,\alpha_{goal}=0$. This case is the reverse of the previous case where the matching section aids in transporting a beam with an extremely small $\beta$ in the PA section to a much larger $\beta$ needed to inject the beam into the undulator section. In Fig. 3(f), we see very good agreement between theory and simulations and that the electron beam emittance is preserved. 
Finally we note that matching of the beam between two  PA sections is essentially combining the cases shown in Figs. 3 (c) and (e). 





In conclusion, we have shown through theory and simulations that exact matching of electron beams into or out of plasma accelerator sections and thereby emittance preservation can be achieved by using longitudinally tailored plasma structures at the entrance or exit of the plasma accelerators while operating in the nonlinear blowout regime.

Work supported by NSFC grants 11175102, 11005063, thousand young talents program, DOE grants DE-SC0010064, DE-SC0008491, DE-SC0008316, and NSF grants ACI-1339893, PHY-1415386, PHY-0960344. Simulations are performed on the UCLA Hoffman 2 Cluster, and Dawson 2 cluster.

\bibliography{refs_xinlu}

\begin{thebibliography}{39}
\expandafter\ifx\csname natexlab\endcsname\relax\def\natexlab#1{#1}\fi
\expandafter\ifx\csname bibnamefont\endcsname\relax
  \def\bibnamefont#1{#1}\fi
\expandafter\ifx\csname bibfnamefont\endcsname\relax
  \def\bibfnamefont#1{#1}\fi
\expandafter\ifx\csname citenamefont\endcsname\relax
  \def\citenamefont#1{#1}\fi
\expandafter\ifx\csname url\endcsname\relax
  \def\url#1{\texttt{#1}}\fi
\expandafter\ifx\csname urlprefix\endcsname\relax\def\urlprefix{URL }\fi
\providecommand{\bibinfo}[2]{#2}
\providecommand{\eprint}[2][]{\url{#2}}

\bibitem[{\citenamefont{Sessler and Wilson}(2007)}]{sessler2007engines}
\bibinfo{author}{\bibfnamefont{A.~M.} \bibnamefont{Sessler}} \bibnamefont{and}
  \bibinfo{author}{\bibfnamefont{E.~J.} \bibnamefont{Wilson}},
  \emph{\bibinfo{title}{Engines of discovery: A century of particle
  accelerators}}, vol.~\bibinfo{volume}{17} (\bibinfo{publisher}{World
  Scientific Hackensack, NJ}, \bibinfo{year}{2007}).

\bibitem[{\citenamefont{Aad et~al.}(2012)\citenamefont{Aad, Abajyan, Abbott
  et~al.}}]{Aad20121}
\bibinfo{author}{\bibfnamefont{G.}~\bibnamefont{Aad}},
  \bibinfo{author}{\bibfnamefont{T.}~\bibnamefont{Abajyan}},
  \bibinfo{author}{\bibfnamefont{B.}~\bibnamefont{Abbott}},
  \bibnamefont{et~al.}, \bibinfo{journal}{Physics Letters B}
  \textbf{\bibinfo{volume}{716}}, \bibinfo{pages}{1 } (\bibinfo{year}{2012}).

\bibitem[{\citenamefont{Emma et~al.}(2010)\citenamefont{Emma, Akre, Arthur,
  Bionta, Bostedt, Bozek, Brachmann, Bucksbaum, Coffee, Decker
  et~al.}}]{emma2010first}
\bibinfo{author}{\bibfnamefont{P.}~\bibnamefont{Emma}},
  \bibinfo{author}{\bibfnamefont{R.}~\bibnamefont{Akre}},
  \bibinfo{author}{\bibfnamefont{J.}~\bibnamefont{Arthur}},
  \bibinfo{author}{\bibfnamefont{R.}~\bibnamefont{Bionta}},
  \bibinfo{author}{\bibfnamefont{C.}~\bibnamefont{Bostedt}},
  \bibinfo{author}{\bibfnamefont{J.}~\bibnamefont{Bozek}},
  \bibinfo{author}{\bibfnamefont{A.}~\bibnamefont{Brachmann}},
  \bibinfo{author}{\bibfnamefont{P.}~\bibnamefont{Bucksbaum}},
  \bibinfo{author}{\bibfnamefont{R.}~\bibnamefont{Coffee}},
  \bibinfo{author}{\bibfnamefont{F.-J.} \bibnamefont{Decker}},
  \bibnamefont{et~al.}, \bibinfo{journal}{nature photonics}
  \textbf{\bibinfo{volume}{4}}, \bibinfo{pages}{641} (\bibinfo{year}{2010}).

\bibitem[{\citenamefont{Litos et~al.}(2014)\citenamefont{Litos, Adli, An,
  Clarke, Clayton, Corde, Delahaye, England, Fisher, Frederico
  et~al.}}]{litos2014high}
\bibinfo{author}{\bibfnamefont{M.}~\bibnamefont{Litos}},
  \bibinfo{author}{\bibfnamefont{E.}~\bibnamefont{Adli}},
  \bibinfo{author}{\bibfnamefont{W.}~\bibnamefont{An}},
  \bibinfo{author}{\bibfnamefont{C.}~\bibnamefont{Clarke}},
  \bibinfo{author}{\bibfnamefont{C.}~\bibnamefont{Clayton}},
  \bibinfo{author}{\bibfnamefont{S.}~\bibnamefont{Corde}},
  \bibinfo{author}{\bibfnamefont{J.}~\bibnamefont{Delahaye}},
  \bibinfo{author}{\bibfnamefont{R.}~\bibnamefont{England}},
  \bibinfo{author}{\bibfnamefont{A.}~\bibnamefont{Fisher}},
  \bibinfo{author}{\bibfnamefont{J.}~\bibnamefont{Frederico}},
  \bibnamefont{et~al.}, \bibinfo{journal}{Nature}
  \textbf{\bibinfo{volume}{515}}, \bibinfo{pages}{92} (\bibinfo{year}{2014}).

\bibitem[{\citenamefont{Leemans et~al.}(2006)}]{leemans2006gev}
\bibinfo{author}{\bibfnamefont{W.}~\bibnamefont{Leemans}} \bibnamefont{et~al.},
  \bibinfo{journal}{Nature Phys.} \textbf{\bibinfo{volume}{2}},
  \bibinfo{pages}{696} (\bibinfo{year}{2006}).

\bibitem[{\citenamefont{Blumenfeld et~al.}(2007)}]{blumenfeld2007energy}
\bibinfo{author}{\bibfnamefont{I.}~\bibnamefont{Blumenfeld}}
  \bibnamefont{et~al.}, \bibinfo{journal}{Nature}
  \textbf{\bibinfo{volume}{445}}, \bibinfo{pages}{741} (\bibinfo{year}{2007}).

\bibitem[{\citenamefont{Kneip et~al.}(2009)\citenamefont{Kneip, Nagel, Martins,
  Mangles, Bellei, Chekhlov, Clarke, Delerue, Divall, Doucas
  et~al.}}]{PhysRevLett.103.035002}
\bibinfo{author}{\bibfnamefont{S.}~\bibnamefont{Kneip}},
  \bibinfo{author}{\bibfnamefont{S.~R.} \bibnamefont{Nagel}},
  \bibinfo{author}{\bibfnamefont{S.~F.} \bibnamefont{Martins}},
  \bibinfo{author}{\bibfnamefont{S.~P.~D.} \bibnamefont{Mangles}},
  \bibinfo{author}{\bibfnamefont{C.}~\bibnamefont{Bellei}},
  \bibinfo{author}{\bibfnamefont{O.}~\bibnamefont{Chekhlov}},
  \bibinfo{author}{\bibfnamefont{R.~J.} \bibnamefont{Clarke}},
  \bibinfo{author}{\bibfnamefont{N.}~\bibnamefont{Delerue}},
  \bibinfo{author}{\bibfnamefont{E.~J.} \bibnamefont{Divall}},
  \bibinfo{author}{\bibfnamefont{G.}~\bibnamefont{Doucas}},
  \bibnamefont{et~al.}, \bibinfo{journal}{Phys. Rev. Lett.}
  \textbf{\bibinfo{volume}{103}}, \bibinfo{pages}{035002}
  (\bibinfo{year}{2009}).

\bibitem[{\citenamefont{Froula et~al.}(2009)\citenamefont{Froula, Clayton,
  D\"oppner, Marsh, Barty, Divol, Fonseca, Glenzer, Joshi, Lu
  et~al.}}]{PhysRevLett.103.215006}
\bibinfo{author}{\bibfnamefont{D.~H.} \bibnamefont{Froula}},
  \bibinfo{author}{\bibfnamefont{C.~E.} \bibnamefont{Clayton}},
  \bibinfo{author}{\bibfnamefont{T.}~\bibnamefont{D\"oppner}},
  \bibinfo{author}{\bibfnamefont{K.~A.} \bibnamefont{Marsh}},
  \bibinfo{author}{\bibfnamefont{C.~P.~J.} \bibnamefont{Barty}},
  \bibinfo{author}{\bibfnamefont{L.}~\bibnamefont{Divol}},
  \bibinfo{author}{\bibfnamefont{R.~A.} \bibnamefont{Fonseca}},
  \bibinfo{author}{\bibfnamefont{S.~H.} \bibnamefont{Glenzer}},
  \bibinfo{author}{\bibfnamefont{C.}~\bibnamefont{Joshi}},
  \bibinfo{author}{\bibfnamefont{W.}~\bibnamefont{Lu}}, \bibnamefont{et~al.},
  \bibinfo{journal}{Phys. Rev. Lett.} \textbf{\bibinfo{volume}{103}},
  \bibinfo{pages}{215006} (\bibinfo{year}{2009}).

\bibitem[{\citenamefont{Clayton et~al.}(2010)\citenamefont{Clayton, Ralph,
  Albert, Fonseca, Glenzer, Joshi, Lu, Marsh, Martins, Mori
  et~al.}}]{PhysRevLett.105.105003}
\bibinfo{author}{\bibfnamefont{C.~E.} \bibnamefont{Clayton}},
  \bibinfo{author}{\bibfnamefont{J.~E.} \bibnamefont{Ralph}},
  \bibinfo{author}{\bibfnamefont{F.}~\bibnamefont{Albert}},
  \bibinfo{author}{\bibfnamefont{R.~A.} \bibnamefont{Fonseca}},
  \bibinfo{author}{\bibfnamefont{S.~H.} \bibnamefont{Glenzer}},
  \bibinfo{author}{\bibfnamefont{C.}~\bibnamefont{Joshi}},
  \bibinfo{author}{\bibfnamefont{W.}~\bibnamefont{Lu}},
  \bibinfo{author}{\bibfnamefont{K.~A.} \bibnamefont{Marsh}},
  \bibinfo{author}{\bibfnamefont{S.~F.} \bibnamefont{Martins}},
  \bibinfo{author}{\bibfnamefont{W.~B.} \bibnamefont{Mori}},
  \bibnamefont{et~al.}, \bibinfo{journal}{Phys. Rev. Lett.}
  \textbf{\bibinfo{volume}{105}}, \bibinfo{pages}{105003}
  (\bibinfo{year}{2010}).

\bibitem[{\citenamefont{Wang et~al.}(2013)}]{wang2013quasi}
\bibinfo{author}{\bibfnamefont{X.}~\bibnamefont{Wang}} \bibnamefont{et~al.},
  \bibinfo{journal}{Nature communications} \textbf{\bibinfo{volume}{4}},
  \bibinfo{pages}{1988} (\bibinfo{year}{2013}).

\bibitem[{\citenamefont{Kim et~al.}(2013)\citenamefont{Kim, Pae, Cha, Kim, Yu,
  Sung, Lee, Jeong, and Lee}}]{PhysRevLett.111.165002}
\bibinfo{author}{\bibfnamefont{H.~T.} \bibnamefont{Kim}},
  \bibinfo{author}{\bibfnamefont{K.~H.} \bibnamefont{Pae}},
  \bibinfo{author}{\bibfnamefont{H.~J.} \bibnamefont{Cha}},
  \bibinfo{author}{\bibfnamefont{I.~J.} \bibnamefont{Kim}},
  \bibinfo{author}{\bibfnamefont{T.~J.} \bibnamefont{Yu}},
  \bibinfo{author}{\bibfnamefont{J.~H.} \bibnamefont{Sung}},
  \bibinfo{author}{\bibfnamefont{S.~K.} \bibnamefont{Lee}},
  \bibinfo{author}{\bibfnamefont{T.~M.} \bibnamefont{Jeong}}, \bibnamefont{and}
  \bibinfo{author}{\bibfnamefont{J.}~\bibnamefont{Lee}},
  \bibinfo{journal}{Phys. Rev. Lett.} \textbf{\bibinfo{volume}{111}},
  \bibinfo{pages}{165002} (\bibinfo{year}{2013}).

\bibitem[{\citenamefont{Leemans et~al.}(2014)\citenamefont{Leemans, Gonsalves,
  Mao, Nakamura, Benedetti, Schroeder, T\'oth, Daniels, Mittelberger, Bulanov
  et~al.}}]{PhysRevLett.113.245002}
\bibinfo{author}{\bibfnamefont{W.~P.} \bibnamefont{Leemans}},
  \bibinfo{author}{\bibfnamefont{A.~J.} \bibnamefont{Gonsalves}},
  \bibinfo{author}{\bibfnamefont{H.-S.} \bibnamefont{Mao}},
  \bibinfo{author}{\bibfnamefont{K.}~\bibnamefont{Nakamura}},
  \bibinfo{author}{\bibfnamefont{C.}~\bibnamefont{Benedetti}},
  \bibinfo{author}{\bibfnamefont{C.~B.} \bibnamefont{Schroeder}},
  \bibinfo{author}{\bibfnamefont{C.}~\bibnamefont{T\'oth}},
  \bibinfo{author}{\bibfnamefont{J.}~\bibnamefont{Daniels}},
  \bibinfo{author}{\bibfnamefont{D.~E.} \bibnamefont{Mittelberger}},
  \bibinfo{author}{\bibfnamefont{S.~S.} \bibnamefont{Bulanov}},
  \bibnamefont{et~al.}, \bibinfo{journal}{Phys. Rev. Lett.}
  \textbf{\bibinfo{volume}{113}}, \bibinfo{pages}{245002}
  (\bibinfo{year}{2014}).

\bibitem[{\citenamefont{Antici et~al.}(2012)\citenamefont{Antici, Bacci,
  Benedetti, Chiadroni, Ferrario, Rossi, Lancia, Migliorati, Mostacci, Palumbo
  et~al.}}]{Antici2012JAP}
\bibinfo{author}{\bibfnamefont{P.}~\bibnamefont{Antici}},
  \bibinfo{author}{\bibfnamefont{A.}~\bibnamefont{Bacci}},
  \bibinfo{author}{\bibfnamefont{C.}~\bibnamefont{Benedetti}},
  \bibinfo{author}{\bibfnamefont{E.}~\bibnamefont{Chiadroni}},
  \bibinfo{author}{\bibfnamefont{M.}~\bibnamefont{Ferrario}},
  \bibinfo{author}{\bibfnamefont{A.~R.} \bibnamefont{Rossi}},
  \bibinfo{author}{\bibfnamefont{L.}~\bibnamefont{Lancia}},
  \bibinfo{author}{\bibfnamefont{M.}~\bibnamefont{Migliorati}},
  \bibinfo{author}{\bibfnamefont{A.}~\bibnamefont{Mostacci}},
  \bibinfo{author}{\bibfnamefont{L.}~\bibnamefont{Palumbo}},
  \bibnamefont{et~al.}, \bibinfo{journal}{Journal of Applied Physics}
  \textbf{\bibinfo{volume}{112}}, \bibinfo{eid}{044902} (\bibinfo{year}{2012}).

\bibitem[{\citenamefont{Migliorati et~al.}(2013)\citenamefont{Migliorati,
  Bacci, Benedetti, Chiadroni, Ferrario, Mostacci, Palumbo, Rossi, Serafini,
  and Antici}}]{PhysRevSTAB.16.011302}
\bibinfo{author}{\bibfnamefont{M.}~\bibnamefont{Migliorati}},
  \bibinfo{author}{\bibfnamefont{A.}~\bibnamefont{Bacci}},
  \bibinfo{author}{\bibfnamefont{C.}~\bibnamefont{Benedetti}},
  \bibinfo{author}{\bibfnamefont{E.}~\bibnamefont{Chiadroni}},
  \bibinfo{author}{\bibfnamefont{M.}~\bibnamefont{Ferrario}},
  \bibinfo{author}{\bibfnamefont{A.}~\bibnamefont{Mostacci}},
  \bibinfo{author}{\bibfnamefont{L.}~\bibnamefont{Palumbo}},
  \bibinfo{author}{\bibfnamefont{A.~R.} \bibnamefont{Rossi}},
  \bibinfo{author}{\bibfnamefont{L.}~\bibnamefont{Serafini}}, \bibnamefont{and}
  \bibinfo{author}{\bibfnamefont{P.}~\bibnamefont{Antici}},
  \bibinfo{journal}{Phys. Rev. ST Accel. Beams} \textbf{\bibinfo{volume}{16}},
  \bibinfo{pages}{011302} (\bibinfo{year}{2013}).

\bibitem[{\citenamefont{Mehrling et~al.}(2012)\citenamefont{Mehrling,
  Grebenyuk, Tsung, Floettmann, and Osterhoff}}]{PhysRevSTAB.15.111303}
\bibinfo{author}{\bibfnamefont{T.}~\bibnamefont{Mehrling}},
  \bibinfo{author}{\bibfnamefont{J.}~\bibnamefont{Grebenyuk}},
  \bibinfo{author}{\bibfnamefont{F.~S.} \bibnamefont{Tsung}},
  \bibinfo{author}{\bibfnamefont{K.}~\bibnamefont{Floettmann}},
  \bibnamefont{and}
  \bibinfo{author}{\bibfnamefont{J.}~\bibnamefont{Osterhoff}},
  \bibinfo{journal}{Phys. Rev. ST Accel. Beams} \textbf{\bibinfo{volume}{15}},
  \bibinfo{pages}{111303} (\bibinfo{year}{2012}).

\bibitem[{\citenamefont{Xu et~al.}(2014)}]{PhysRevLett.112.035003}
\bibinfo{author}{\bibfnamefont{X.~L.} \bibnamefont{Xu}} \bibnamefont{et~al.},
  \bibinfo{journal}{Phys. Rev. Lett.} \textbf{\bibinfo{volume}{112}},
  \bibinfo{pages}{035003} (\bibinfo{year}{2014}).

\bibitem[{\citenamefont{Floettmann}(2014)}]{PhysRevSTAB.17.054402}
\bibinfo{author}{\bibfnamefont{K.}~\bibnamefont{Floettmann}},
  \bibinfo{journal}{Phys. Rev. ST Accel. Beams} \textbf{\bibinfo{volume}{17}},
  \bibinfo{pages}{054402} (\bibinfo{year}{2014}).

\bibitem[{\citenamefont{Dornmair et~al.}(2015)}]{PhysRevSTAB.18.041302}
\bibinfo{author}{\bibfnamefont{I.}~\bibnamefont{Dornmair}}
  \bibnamefont{et~al.}, \bibinfo{journal}{Phys. Rev. ST Accel. Beams}
  \textbf{\bibinfo{volume}{18}}, \bibinfo{pages}{041302}
  (\bibinfo{year}{2015}).

\bibitem[{\citenamefont{Fonseca et~al.}(2002)}]{fonseca2002high}
\bibinfo{author}{\bibfnamefont{R.}~\bibnamefont{Fonseca}} \bibnamefont{et~al.},
  \bibinfo{journal}{Lecture notes in computer science}
  \textbf{\bibinfo{volume}{2331}}, \bibinfo{pages}{342} (\bibinfo{year}{2002}).

\bibitem[{\citenamefont{Rosenzweig et~al.}(1991)\citenamefont{Rosenzweig,
  Breizman, Katsouleas, and Su}}]{PhysRevA.44.R6189}
\bibinfo{author}{\bibfnamefont{J.~B.} \bibnamefont{Rosenzweig}},
  \bibinfo{author}{\bibfnamefont{B.}~\bibnamefont{Breizman}},
  \bibinfo{author}{\bibfnamefont{T.}~\bibnamefont{Katsouleas}},
  \bibnamefont{and} \bibinfo{author}{\bibfnamefont{J.~J.} \bibnamefont{Su}},
  \bibinfo{journal}{Phys. Rev. A} \textbf{\bibinfo{volume}{44}},
  \bibinfo{pages}{R6189} (\bibinfo{year}{1991}).

\bibitem[{\citenamefont{Lu et~al.}(2006{\natexlab{a}})}]{PhysRevLett.96.165002}
\bibinfo{author}{\bibfnamefont{W.}~\bibnamefont{Lu}} \bibnamefont{et~al.},
  \bibinfo{journal}{Phys. Rev. Lett.} \textbf{\bibinfo{volume}{96}},
  \bibinfo{pages}{165002} (\bibinfo{year}{2006}{\natexlab{a}}).

\bibitem[{\citenamefont{Lu et~al.}(2006{\natexlab{b}})}]{lu2006nonlinearPoP}
\bibinfo{author}{\bibfnamefont{W.}~\bibnamefont{Lu}} \bibnamefont{et~al.},
  \bibinfo{journal}{Phys. Plasma} \textbf{\bibinfo{volume}{13}},
  \bibinfo{pages}{056709} (\bibinfo{year}{2006}{\natexlab{b}}).

\bibitem[{\citenamefont{Lee}(1999)}]{lee1999accelerator}
\bibinfo{author}{\bibfnamefont{S.-Y.} \bibnamefont{Lee}},
  \emph{\bibinfo{title}{Accelerator physics}} (\bibinfo{publisher}{World
  Scientific Singapore}, \bibinfo{year}{1999}).

\bibitem[{\citenamefont{Gonsalves et~al.}(2011)\citenamefont{Gonsalves,
  Nakamura, Lin, Panasenko, Shiraishi, Sokollik, Benedetti, Schroeder, Geddes,
  Van~Tilborg et~al.}}]{gonsalves2011tunable}
\bibinfo{author}{\bibfnamefont{A.}~\bibnamefont{Gonsalves}},
  \bibinfo{author}{\bibfnamefont{K.}~\bibnamefont{Nakamura}},
  \bibinfo{author}{\bibfnamefont{C.}~\bibnamefont{Lin}},
  \bibinfo{author}{\bibfnamefont{D.}~\bibnamefont{Panasenko}},
  \bibinfo{author}{\bibfnamefont{S.}~\bibnamefont{Shiraishi}},
  \bibinfo{author}{\bibfnamefont{T.}~\bibnamefont{Sokollik}},
  \bibinfo{author}{\bibfnamefont{C.}~\bibnamefont{Benedetti}},
  \bibinfo{author}{\bibfnamefont{C.}~\bibnamefont{Schroeder}},
  \bibinfo{author}{\bibfnamefont{C.}~\bibnamefont{Geddes}},
  \bibinfo{author}{\bibfnamefont{J.}~\bibnamefont{Van~Tilborg}},
  \bibnamefont{et~al.}, \bibinfo{journal}{Nature Physics}
  \textbf{\bibinfo{volume}{7}}, \bibinfo{pages}{862} (\bibinfo{year}{2011}).

\bibitem[{\citenamefont{Liu et~al.}(2011)\citenamefont{Liu, Xia, Wang, Lu,
  Wang, Deng, Li, Zhang, Liang, Leng et~al.}}]{PhysRevLett.107.035001}
\bibinfo{author}{\bibfnamefont{J.~S.} \bibnamefont{Liu}},
  \bibinfo{author}{\bibfnamefont{C.~Q.} \bibnamefont{Xia}},
  \bibinfo{author}{\bibfnamefont{W.~T.} \bibnamefont{Wang}},
  \bibinfo{author}{\bibfnamefont{H.~Y.} \bibnamefont{Lu}},
  \bibinfo{author}{\bibfnamefont{C.}~\bibnamefont{Wang}},
  \bibinfo{author}{\bibfnamefont{A.~H.} \bibnamefont{Deng}},
  \bibinfo{author}{\bibfnamefont{W.~T.} \bibnamefont{Li}},
  \bibinfo{author}{\bibfnamefont{H.}~\bibnamefont{Zhang}},
  \bibinfo{author}{\bibfnamefont{X.~Y.} \bibnamefont{Liang}},
  \bibinfo{author}{\bibfnamefont{Y.~X.} \bibnamefont{Leng}},
  \bibnamefont{et~al.}, \bibinfo{journal}{Phys. Rev. Lett.}
  \textbf{\bibinfo{volume}{107}}, \bibinfo{pages}{035001}
  (\bibinfo{year}{2011}).

\bibitem[{\citenamefont{Clayton et~al.}(1993)\citenamefont{Clayton, Marsh,
  Dyson, Everett, Lal, Leemans, Williams, and Joshi}}]{PhysRevLett.70.37}
\bibinfo{author}{\bibfnamefont{C.~E.} \bibnamefont{Clayton}},
  \bibinfo{author}{\bibfnamefont{K.~A.} \bibnamefont{Marsh}},
  \bibinfo{author}{\bibfnamefont{A.}~\bibnamefont{Dyson}},
  \bibinfo{author}{\bibfnamefont{M.}~\bibnamefont{Everett}},
  \bibinfo{author}{\bibfnamefont{A.}~\bibnamefont{Lal}},
  \bibinfo{author}{\bibfnamefont{W.~P.} \bibnamefont{Leemans}},
  \bibinfo{author}{\bibfnamefont{R.}~\bibnamefont{Williams}}, \bibnamefont{and}
  \bibinfo{author}{\bibfnamefont{C.}~\bibnamefont{Joshi}},
  \bibinfo{journal}{Phys. Rev. Lett.} \textbf{\bibinfo{volume}{70}},
  \bibinfo{pages}{37} (\bibinfo{year}{1993}).

\bibitem[{\citenamefont{Clayton et~al.}(1994)\citenamefont{Clayton, Everett,
  Lal, Gordon, Marsh, and Joshi}}]{clayton1994acceleration}
\bibinfo{author}{\bibfnamefont{C.}~\bibnamefont{Clayton}},
  \bibinfo{author}{\bibfnamefont{M.}~\bibnamefont{Everett}},
  \bibinfo{author}{\bibfnamefont{A.}~\bibnamefont{Lal}},
  \bibinfo{author}{\bibfnamefont{D.}~\bibnamefont{Gordon}},
  \bibinfo{author}{\bibfnamefont{K.}~\bibnamefont{Marsh}}, \bibnamefont{and}
  \bibinfo{author}{\bibfnamefont{C.}~\bibnamefont{Joshi}},
  \bibinfo{journal}{Physics of Plasmas (1994-present)}
  \textbf{\bibinfo{volume}{1}}, \bibinfo{pages}{1753} (\bibinfo{year}{1994}).

\bibitem[{\citenamefont{Everett et~al.}(1994)\citenamefont{Everett, Lal,
  Gordon, Clayton, Marsh, and Joshi}}]{everett1994trapped}
\bibinfo{author}{\bibfnamefont{M.}~\bibnamefont{Everett}},
  \bibinfo{author}{\bibfnamefont{A.}~\bibnamefont{Lal}},
  \bibinfo{author}{\bibfnamefont{D.}~\bibnamefont{Gordon}},
  \bibinfo{author}{\bibfnamefont{C.}~\bibnamefont{Clayton}},
  \bibinfo{author}{\bibfnamefont{K.}~\bibnamefont{Marsh}}, \bibnamefont{and}
  \bibinfo{author}{\bibfnamefont{C.}~\bibnamefont{Joshi}},
  \bibinfo{journal}{Nature} \textbf{\bibinfo{volume}{368}},
  \bibinfo{pages}{527} (\bibinfo{year}{1994}).

\bibitem[{\citenamefont{Stragier et~al.}(2011)\citenamefont{Stragier, Luiten,
  van~der Geer, van~der Wiel, and Brussaard}}]{Stragier2011externalinjection}
\bibinfo{author}{\bibfnamefont{X.~F.~D.} \bibnamefont{Stragier}},
  \bibinfo{author}{\bibfnamefont{O.~J.} \bibnamefont{Luiten}},
  \bibinfo{author}{\bibfnamefont{S.~B.} \bibnamefont{van~der Geer}},
  \bibinfo{author}{\bibfnamefont{M.~J.} \bibnamefont{van~der Wiel}},
  \bibnamefont{and} \bibinfo{author}{\bibfnamefont{G.~J.~H.}
  \bibnamefont{Brussaard}}, \bibinfo{journal}{Journal of Applied Physics}
  \textbf{\bibinfo{volume}{110}}, \bibinfo{eid}{024910} (\bibinfo{year}{2011}).

\bibitem[{\citenamefont{Rossi et~al.}(2014)\citenamefont{Rossi, Bacci,
  Belleveglia, Chiadroni, Cianchi, Pirro, Ferrario, Gallo, Gatti, Maroli
  et~al.}}]{Rossi201460}
\bibinfo{author}{\bibfnamefont{A.~R.} \bibnamefont{Rossi}},
  \bibinfo{author}{\bibfnamefont{A.}~\bibnamefont{Bacci}},
  \bibinfo{author}{\bibfnamefont{M.}~\bibnamefont{Belleveglia}},
  \bibinfo{author}{\bibfnamefont{E.}~\bibnamefont{Chiadroni}},
  \bibinfo{author}{\bibfnamefont{A.}~\bibnamefont{Cianchi}},
  \bibinfo{author}{\bibfnamefont{G.~D.} \bibnamefont{Pirro}},
  \bibinfo{author}{\bibfnamefont{M.}~\bibnamefont{Ferrario}},
  \bibinfo{author}{\bibfnamefont{A.}~\bibnamefont{Gallo}},
  \bibinfo{author}{\bibfnamefont{G.}~\bibnamefont{Gatti}},
  \bibinfo{author}{\bibfnamefont{C.}~\bibnamefont{Maroli}},
  \bibnamefont{et~al.}, \bibinfo{journal}{Nucl. Instr. and Meth. A}
  \textbf{\bibinfo{volume}{740}}, \bibinfo{pages}{60 } (\bibinfo{year}{2014}).

\bibitem[{\citenamefont{Schlenvoigt et~al.}(2007)}]{schlenvoigt2007compact}
\bibinfo{author}{\bibfnamefont{H.-P.} \bibnamefont{Schlenvoigt}}
  \bibnamefont{et~al.}, \bibinfo{journal}{Nature Physics}
  \textbf{\bibinfo{volume}{4}}, \bibinfo{pages}{130} (\bibinfo{year}{2007}).

\bibitem[{\citenamefont{Fuchs et~al.}(2009)}]{fuchs2009laser}
\bibinfo{author}{\bibfnamefont{M.}~\bibnamefont{Fuchs}} \bibnamefont{et~al.},
  \bibinfo{journal}{Nature physics} \textbf{\bibinfo{volume}{5}},
  \bibinfo{pages}{826} (\bibinfo{year}{2009}).

\bibitem[{\citenamefont{Cipiccia et~al.}(2011)}]{cipiccia2011gamma}
\bibinfo{author}{\bibfnamefont{S.}~\bibnamefont{Cipiccia}}
  \bibnamefont{et~al.}, \bibinfo{journal}{Nature Physics}
  \textbf{\bibinfo{volume}{7}}, \bibinfo{pages}{867} (\bibinfo{year}{2011}).

\bibitem[{\citenamefont{Leemans and Esarey}(2009)}]{leemans2009laser}
\bibinfo{author}{\bibfnamefont{W.}~\bibnamefont{Leemans}} \bibnamefont{and}
  \bibinfo{author}{\bibfnamefont{E.}~\bibnamefont{Esarey}},
  \bibinfo{journal}{Physics Today} \textbf{\bibinfo{volume}{62}},
  \bibinfo{pages}{44} (\bibinfo{year}{2009}).

\bibitem[{\citenamefont{Adli et~al.}(2013)\citenamefont{Adli, Delahaye,
  Gessner, Hogan, Raubenheimer, An, Joshi, and Mori}}]{adli2013beam}
\bibinfo{author}{\bibfnamefont{E.}~\bibnamefont{Adli}},
  \bibinfo{author}{\bibfnamefont{J.-P.} \bibnamefont{Delahaye}},
  \bibinfo{author}{\bibfnamefont{S.~J.} \bibnamefont{Gessner}},
  \bibinfo{author}{\bibfnamefont{M.~J.} \bibnamefont{Hogan}},
  \bibinfo{author}{\bibfnamefont{T.}~\bibnamefont{Raubenheimer}},
  \bibinfo{author}{\bibfnamefont{W.}~\bibnamefont{An}},
  \bibinfo{author}{\bibfnamefont{C.}~\bibnamefont{Joshi}}, \bibnamefont{and}
  \bibinfo{author}{\bibfnamefont{W.}~\bibnamefont{Mori}},
  \bibinfo{journal}{arXiv preprint arXiv:1308.1145}  (\bibinfo{year}{2013}).

\bibitem[{\citenamefont{Eichner et~al.}(2007)\citenamefont{Eichner, Gr\"uner,
  Becker, Fuchs, Habs, Weingartner, Schramm, Backe, Kunz, and
  Lauth}}]{PhysRevSTAB.10.082401}
\bibinfo{author}{\bibfnamefont{T.}~\bibnamefont{Eichner}},
  \bibinfo{author}{\bibfnamefont{F.}~\bibnamefont{Gr\"uner}},
  \bibinfo{author}{\bibfnamefont{S.}~\bibnamefont{Becker}},
  \bibinfo{author}{\bibfnamefont{M.}~\bibnamefont{Fuchs}},
  \bibinfo{author}{\bibfnamefont{D.}~\bibnamefont{Habs}},
  \bibinfo{author}{\bibfnamefont{R.}~\bibnamefont{Weingartner}},
  \bibinfo{author}{\bibfnamefont{U.}~\bibnamefont{Schramm}},
  \bibinfo{author}{\bibfnamefont{H.}~\bibnamefont{Backe}},
  \bibinfo{author}{\bibfnamefont{P.}~\bibnamefont{Kunz}}, \bibnamefont{and}
  \bibinfo{author}{\bibfnamefont{W.}~\bibnamefont{Lauth}},
  \bibinfo{journal}{Phys. Rev. ST Accel. Beams} \textbf{\bibinfo{volume}{10}},
  \bibinfo{pages}{082401} (\bibinfo{year}{2007}).

\bibitem[{\citenamefont{Harrison et~al.}(2012)\citenamefont{Harrison, Joshi,
  Lake, Candler, and Musumeci}}]{PhysRevSTAB.15.070703}
\bibinfo{author}{\bibfnamefont{J.}~\bibnamefont{Harrison}},
  \bibinfo{author}{\bibfnamefont{A.}~\bibnamefont{Joshi}},
  \bibinfo{author}{\bibfnamefont{J.}~\bibnamefont{Lake}},
  \bibinfo{author}{\bibfnamefont{R.}~\bibnamefont{Candler}}, \bibnamefont{and}
  \bibinfo{author}{\bibfnamefont{P.}~\bibnamefont{Musumeci}},
  \bibinfo{journal}{Phys. Rev. ST Accel. Beams} \textbf{\bibinfo{volume}{15}},
  \bibinfo{pages}{070703} (\bibinfo{year}{2012}).

\bibitem[{\citenamefont{Tzoufras et~al.}(2008)}]{PhysRevLett.101.145002}
\bibinfo{author}{\bibfnamefont{M.}~\bibnamefont{Tzoufras}}
  \bibnamefont{et~al.}, \bibinfo{journal}{Phys. Rev. Lett.}
  \textbf{\bibinfo{volume}{101}}, \bibinfo{pages}{145002}
  (\bibinfo{year}{2008}).

\bibitem[{mat()}]{mathematica}
\bibinfo{note}{Mathematica, version nine, 2012,
  http://www.wolfram.com/mathematica/.}

\end{thebibliography}

\end{document}